\documentclass[
aps,%
10pt,%
final,%
notitlepage,%
oneside,
twocolumn,%
nobibnotes,%
nofootinbib,%
superscriptaddress,%
noshowpacs,%
centertags]%
{revtex4}

\begin{document}
\title{Radiation Pattern of Radio and Optical Components
 of Extended Radio Sources}

\author{V.R.~Amirkhanyan}
 \affiliation{Special Astrophysical Observatory, Russian Academy of Sciences, Nizhnij Arkhyz, 369167, Russia}
 \affiliation{Sternberg Astronomical Institute, M.V.Lomonosov Moscow State University, Moscow, 119992 Russia}
 \email{amir@sao.ru}

\begin{abstract}
The relation between parameters the $D/\sqrt{I}$ and $I_C/I_{\rm SUM}$
 and radiation patterns of the optical and radio components
of an extended radio source is analyzed, where $D$ and $I$ are the
apparent size and observed radiation intensity of the
source or its components respectively. The parameters of the pattern in the
optical and radio (1.4~GHz)  ranges are estimated. The radiation
pattern of extended radio-emitting regions is close to spherical and
the radiation of the central component is concentrated in a $24\degr$
wide beam. Its luminosity is a factor of  4.58 higher than that of
the extended component of the radio source. The radiation pattern of
the optical component of the  radio source turned out to be
unexpectedly non-spherical: the main lobe of the pattern is about
$26\degr$ wide. The  $g$-band luminosity is  6.4--12.3~times higher
than the luminosity of the spherical fraction  of the  ``optical''
radiation pattern. A list of 116 new giant radio sources is
presented.
\end{abstract}

\maketitle

\section{INTRODUCTION}
Currently, the radiation pattern (RP) of a classical extended source
is represented by the sum of two models of the radio sources.
\begin{list}{}{
\setlength\leftmargin{2mm} \setlength\topsep{2mm}
\setlength\parsep{0mm} \setlength\itemsep{2mm} }
  \item (1) The Shklovsky--van~der~Laan model~[1],[2]%\cite{1:Amirkhanyan_n_en,2:Amirkhanyan_n_en}
  considers the motion of relativistic electrons in irregular magnetic fields. The radiation pattern of such a radio source is close to spherical.
 \item (2) The model of the radiation of relativistic electrons moving in
 regular magnetic fields~\cite{3:Amirkhanyan_n_en,4:Amirkhanyan_n_en,5:Amirkhanyan_n_en,6:Amirkhanyan_n_en}. The radiation pattern of this structure has the
 form of a narrow beam, which in the simplest case of uniform magnetic field
 can described by the following equation~\cite{7:Amirkhanyan_n_en}
$$
A(\varphi)=[\gamma(1-\beta)\cos\varphi]^{-(2-\alpha)},
$$
where $\varphi$ is the angle between the observer's line of sight and
the direction of the relativistic jet; \mbox
{$\gamma\!=\!{(1-\beta^2)}^{-1/2}$} and
 $\alpha$ is the spectral index \linebreak \mbox {($S_{\nu} \sim\nu^{+\alpha}$)}.
The model yields a simple formula for estimating the of the radiation
pattern
 \mbox{$\Delta\phi=1$/$\gamma$}, which is independent of frequency.
The  $\gamma$ values inferred from observations of superluminal
motions in radio-source jets lie in  the  \mbox {3--40} interval,
which corresponds to $\Delta\phi$ values from $20\degr$ to $1\fdg 5$.
May be that the resulting RP of the central component is equal to
the sum of the patterns of different portions of the jet with
different generation conditions. This conclusion is suggested by the
study of the central region of Centaurus~A radio source by
Horiuchi~et~al.~\cite{8:Amirkhanyan_n_en}, which demonstrates the
variation of the jet opening angle from  12 to $3\degr$ at the distance of
1000 to 20\,000 ~Schwartzschield radii from  the central engine. 
This result suggests that
electrons ejected by the central engine into a wide opening angle
``converge'' as they move away from the center. 
Hence the ``hedgehog'' model with radial magnetic
field whose RP was analyzed by Kovalev~\cite{9:Amirkhanyan_n_en}
applies near the central engine. In this article author performs
computations over a wide frequency range, which imply that the width
 the RP increases with increasing frequency. For example, 1.4~GHz
and 5~GHz $\Delta\phi=13\fdg 3$  and $40\fdg 4$ at 1.4 and 5~GHz,
respectively. Possibly , the truth is in a combination of  models~\cite{7:Amirkhanyan_n_en}
 and~\cite{9:Amirkhanyan_n_en} as that is demanded by dialectic.
 \end{list}

There are very few experimental estimates of the RP of extended radio
sources. This include, first,  the  paper by  Orr and
Browne~\cite{10:Amirkhanyan_n_en}. The above authors combined the RP
from~\cite{7:Amirkhanyan_n_en} with the flux density ratio of the
compact and extended components (parameter \mbox{$R=S_C/S_E$}) and
used it as the indicator of the orientation of the radio source with
respect to the observer's line of sight. They further assumed
equiprobable distribution of the space orientation to construct the
computed distribution of parameter  $R$ and compare it with the
histogram of observed $R$ values for quasars. As a result, they
estimated at 5~GHz the Lorentz factor  $\gamma\sim 5$ and found the
radiation level of the spherical component to be equal Such
parameters imply a RP width of $\Delta\phi=14\degr$.
Amirkhanyan~\cite{11:Amirkhanyan_n_en} used a similar method for
estimating the RP, but unlike Orr and
Browne~\cite{10:Amirkhanyan_n_en} he investigated the statistics of
the ratios of the compact component flux density to the total radio
source flux density, $R_S=S_C/S_{\rm SUM}$. In addition, an upper
redshift of the objects list was limited to prevent selection of the
radio sources orientation with respect to the observer. As a result,
the RP width was estimated to be  \mbox {$\Delta\phi=42\degr$} and
$\Delta\phi=54\degr$ at 1.4 and 5~Ghz, respectively. Yet another
estimate was obtained in~\cite{12:Amirkhanyan_n_en} by constructing
the  ``angular size--redshift''
 dependence, which yielded  $\Delta\phi=24\degr$ at 1.4~GHz.
Complete disorder and reel.

\section{OBSERVED PARAMETERS AND SPATIAL ORIENTATION OF RADIO SOURCES}

To study the RP of an antenna, we change its orientation and measure its radiation.
We cannot similarly manipulate with radio sources. However, we can assume that
the distribution of their orientations is close to equiprobable and try
to find a relation between several parameters that depend on the orientation of
the radio source. Such bonds should inevitably show up if the structure of the
radio source and its RP are interconnected. We assume that this condition
is fulfilled.

 Our conviction that \mbox{$R_S=S_C/S_{\rm SUM}$} can be used as an orientation
 indicator is based on the assumption that RP is not spherical. The statistics
 of \mbox{$R_S$} from whose behavior we conclude about the form of the
 radiation pattern depends very strongly on the sample  of radio sources,
radio-telescope parameters, and eye of the observer. We therefore
need yet another indicator, which inevitably depends on the
parameters of the RP and orientation of the object and which can be
inferred from observational data. We can then compare the behavior of
these parameters with our theoretical studies ,try to find their
connection and understand to which extent our assumptions are
consistent with reality. To approach the truth as close as possible,
we must try to  ``tear off'' connection of our indicators from the space model.

As the first orientation indicator we use the ratio of the emission
of the compact component to the total radiation intensity of the
source \mbox{$R_L = I_C/I_{\rm SUM}$}. Note that by radiation
intensity we mean the intensity in the direction of the observer. In
the general case of non-spherical RP $R_L$ is not equal to the ratio
of the luminosities of the objects.

Transition to  $R_L$ allows us to eliminate the redshift dependence of the supposed
orientation indicator, whereas such dependence is inevitable for
 \mbox{$R_S=S_C/S_{\rm SUM}$}: spectral indices $S_C$ and $S_{\rm SUM}$, and
 hence their $K$ corrections differ from each other.

Let us derive the formula for  $R_L$ in terms of the observed parameters of
extended radio sources. Let the flux density depend on frequency
as $S\propto\nu^\alpha$. Given that \mbox{$I=
Sl_b^2(z)(1+z)^{-(1+\alpha)}$,} where $l_b$ is the bolometric distance,
we write
\smallskip
\\
$R_L= \frac{I_C}{I_{SUM}}=$\\%[+15pt]
$=\frac{S_Cl_b^2(z)(1+z)^{-(1+\alpha_C)}}{\displaystyle S_Cl_b^2(z)(1+z)^{-(1+\alpha_C)}+S_El_b^2(z)(1+z)^{-(1+\alpha_E)}}$.
\smallskip

Here $\alpha_E$ and  $S_E$ are the spectral index and flux density
 of the extended component, respectively, and $\alpha_C$ and  $S_C$ are the spectral index
 and flux density of the compact component,  respectively.
Given that  \mbox{$S_E=S_{\rm SUM}-S_C$}, we derive, after simple transformations,
the following formula

\begin{equation}
\begin{array}{rcl}
R_L&=& \frac{S_C}{S_C +S_{\rm SUM}-S_C(1+z)^{\alpha_C-\alpha_E}}\\[+15pt]
&=&\frac{(1+z)^{\alpha_E-\alpha_C}}{\displaystyle
(1+z)^{\alpha_E-\alpha_C}+\frac{1}{R_S}-1}.
\end{array}
\end{equation}

The second parameter
\begin{equation}
T=\frac{D}{\sqrt{I}},
\end{equation}
 where $D=D_0 \sin\varphi $ is the visible size of the radio source and
$I=I_0\,A(\varphi)$, the ~visible radiation of the radio source or
its component. The word ``visible'' is used to indicate that both $D$
and $I$ depend on the orientation of the radio source relative to the
observer; $\varphi$ is the angle between the line of sight and the
direction toward the maximum of emission of the radio source;
 $D_0$ is the true size of the radio source and  $I_0$ is the radiation
 in the direction of the maximum of the radiation pattern $A(\varphi)$.
 The $T$ value can be  easily computed from the observed parameters
\begin{equation}
T=\frac{\Theta}{\sqrt S}(1+z)^{-\frac{(3-\alpha)}{2}}\\
%T=\frac{\Theta}{\sqrt S}(1+z)^{-\frac{(\vphantom{\int}
%3-\alpha)}{\vphantom{\sqrt{a}} 2}}
\end{equation}
and does not depend on the model of space. $\Theta$ is the angular
size of the radio source and $S$, the flux density of the radio
source or of its component.

The author assumes that substituting the flux density of the entire
radio source or some of its components (compact, extended, optical)
into the denominator in formula~(3) will make it possible to estimate
the parameters of the RP of both the entire radio source and its
corresponding fractions.

Let us relate  $R_L$  and $T$  via RP $A(\varphi)$. Consider the canonical model
of a radio source with symmetric two-sided components~\cite{7:Amirkhanyan_n_en,10:Amirkhanyan_n_en}. Let the ideal
RP of such object be axisymmetric with respect to the jet axis and have two symmetric
maxima in opposite directions. Let us adopt the description of the radiation pattern from~\cite{11:Amirkhanyan_n_en}
\begin{equation}
A(\varphi)=\frac{I}{I_0}=a+(1-a)\cos^{2n}\varphi.
\end{equation}
Here $a$ is the level  of the spherical component of the RP and $n$,
a parameter that determines the width of the main lobe of the RP. If
\mbox{$n$ = 0} then the RP degenerates into a spherical pattern
($A(\varphi)=1$).

Let us now determine the relation between  $R_L$ and the orientation of the
radio source relative to the observer
\begin{equation}
R_L=\frac{I_C}{I_{\rm
SUM}}=\displaystyle{\frac{(1-a)\cos^{2n}\varphi}{a+(1-a)\cos^{2n}\varphi}}.
\end{equation}
It is evident from this equation  that $R_L < 1-a$.

Let us now extract the following relation from formula~(5)
$$\cos\varphi=\left[\displaystyle\frac{aR_L}{(1-a)(1-R_L)}\right]^{\frac{1}{2n}}$$
and, given that $D = D_0 \sin\varphi$, substitute it into
formula~(2). As a result, we obtain the following computed dependence
of  $T$ on $R_L$ for different components of the radio source:

\vspace{2mm}
(1) The total radiation of the  radio source
\begin{equation}
\begin{array}{rcl}
%\begin{split}
T_{\rm SUM}&=&\frac{D}{\sqrt I_{\rm SUM}}=\frac{D_0
\sin\varphi}{\sqrt{I_0[a+(1-a)\cos^{2n}\varphi]}}\\[+15pt]
&=&\frac{D_0\sqrt{1-\left[\displaystyle\frac{aR_L}{(1-a)(1-R_L)}\right]^{\frac{1}{n}}}}{\sqrt{\displaystyle\frac{aI_0}{1-R_L}}};

\end{array}
\end{equation}

(2) The compact component of the radio source
\begin{equation}
\begin{array}{rcl}

T_C&=&\frac{D}{\sqrt I_C}=
\frac{D_0\sin\varphi}{\sqrt{I_0(1-a)\cos^{2n}\varphi}}\\[+15pt]
&=&\frac{D_0\sqrt{1-\left[\displaystyle\frac{aR_L}{(1-a)(1-R_L)}\right]^{\frac{1}{n}}}}{\sqrt{I_0
\displaystyle\frac{aR_L}{1-R_L}}};
\end{array}
\end{equation}
(3) the extended component of the radio source
\begin{equation}
\begin{array}{rcl}
%\begin{split}
T_E&=&\frac{D}{\sqrt I_E}= \frac{D_0 \sin\varphi}{\sqrt{aI_0}}\\[+15pt]
&=&\frac{D_0\sqrt{1-\left[\displaystyle\frac{aR_L}{(1-a)(1-R_L)}\right]^{\frac{1}{n}}}}{\sqrt{aI_0}}.
\end{array}
\end{equation}

Formulas~(6)--(8) include parameters of the RP and therefore there is
hope that these parameters can be estimated by comparing the model
with experiment. Note that all the three equations must demonstrate
the agreement with the experiment for the same parameters of RP.

The optical object is a component of the radio source and we can
extend the above reasoning to it in order to estimate the RP in the
optical range. We now consider the general case allowing
non-spherical radiation of the optical component. We further assume
that the optical radiation is associated with the structure of the
radio source: the symmetry  axis of the optical radiation coincides
with the jet axis. It follows from the canonical model of the radio
source, which assume the synchrotron mechanism of jet radiation over
a wide range of wavelengths and the closeness of the spatial
orientation of the rotation axes of the host galaxy and the jet. Here
we can draw upon the works by
Condon~et~al.~\cite{13:Amirkhanyan_n_en} and Browne and
Battye~\cite{14:Amirkhanyan_n_en}, who  showed based on experimental
data that the the distribution function of the difference of the position 
angles  radio sources and of elliptical galaxies identified with 
them is has maximum  near $90\degr$.
Hence the position angle of the  minor axis, which is close to the
position angle of the normal to the galaxy, coincides with the
orientation of the radio jet. Let the form of the optical RP be
described by formula~(4), but with its proper values of $n_{\rm opt}$
and $a_{\rm opt}$:
\begin{equation}
A_{\rm opt}(\varphi)\!=\!\frac{I_{\rm opt}}{I_{\rm
0,\,opt}}\!=\!a_{\rm opt}\!+\!(1\!-\!a_{\rm opt})\cos^{2n_{\rm
opt}}\varphi\nonumber  ~(4')
\end{equation}

Currently, we cannot separate the radiation  into the extended and 
compact fractions of the optical object. We therefore use
formula~(6) for the total radiation, where we leave parameter $R_L$
determined from radio data as the orientation indicator, but replace
RP~(4) by formula~($4'$). We  then have
\begin{equation}
\begin{array}{l}
T_{\rm opt}\!=\!\frac{D}{\sqrt I_{\rm opt}}\!=\!
\displaystyle\frac{D_0\sin\varphi}{\sqrt{I_{\rm 0,\,opt}[a_{\rm
opt}\!+\!(1\!-\!a_{\rm opt})\cos^{2n_{\rm opt}}\varphi]}}\\[20pt]
\!=\!\frac{D_0\sqrt{1\!-\!\left[\displaystyle\frac{aR_L}{(1\!-\!a)(1\!-\!R_L)}\right]^{\frac{1}{n}}}}{\sqrt{I_{\rm
0,\,opt}\left[a_{\rm opt}\!+\!(1\!-\!a_{\rm
opt})\displaystyle\frac{aR_L}{(1\!-\!a)(1\!-\!R_L)}\right]^{\frac{n_{\rm
opt}}{n}}}}.
\end{array}
\end{equation}

\section{SAMPLES OF RADIO SOURCES}

To compare formulas~(6)--(9) with real measurements, we used two
lists of extended radio sources:
\begin{list}{}{
\setlength\leftmargin{2mm} \setlength\topsep{2mm}
\setlength\parsep{0mm} \setlength\itemsep{2mm} } \item (1) Sample of
objects whose apparent size does not exceed 0.7~Mpc in the $\Lambda
CDM$ model. This list includes  2947 identified radio sources from
Amirkhanyan~\cite{15:Amirkhanyan_n_en}. For each of these objects its
angular size, total flux density, component flux densities, redshift,
and $g$-band magnitude of the optical counterpart are known. \item
(2) Objects whose apparent sizes exceed 0.7~Mpc. This sample contains
254 giant radio sources from catalogs~\cite{15:Amirkhanyan_n_en,
16:Amirkhanyan_n_en,17:Amirkhanyan_n_en, 18:Amirkhanyan_n_en,
19:Amirkhanyan_n_en, 20:Amirkhanyan_n_en, 21:Amirkhanyan_n_en,
22:Amirkhanyan_n_en, 23:Amirkhanyan_n_en, 24:Amirkhanyan_n_en,
25:Amirkhanyan_n_en, 26:Amirkhanyan_n_en}.
\end{list}
The list does not include objects with photometric redshifts. To
ensure uniformity of parameter determination, we found all objects in
the NVSS ({\tt http://www.cv.nrao.edu/nvss/NVSSlist.shtml}) or SUMSS
({\tt
http://www.astrop.physics.usyd.\\edu.au/sumsscat/sumsscat.Feb-16-2012})
catalogs. We thereby determined the flux densities and coordinates of
the radio-source components, and also their angular sizes measured as
the separation between the most spaced components.

If there was a radio component within $10$--$15\arcsec$ of the
optical component then this radio component was considered to be the
central one and $S_C$ was set equal to its flux density. If no
central component was present then its flux density was assigned as
\mbox{$S_C$ = 0.1}~mJy, which is below the detection threshold for
NVSS. The data of the FIRST survey~({\tt
http://sundog.stsci.edu/cgi-bin/\\searchfirst}) were used to refine
the coordinates of the central components if the radio source was
located in the area covered by this survey.

We further added to the list 92 radio sources found in the NVSS
catalog using the software developed by the
author~\cite{26:Amirkhanyan_n_en}~(see Table below). This table also
includes the ``giants''  from~\cite{15:Amirkhanyan_n_en}. The author
broke selection criteria for the radio source 2057\,+\,0012, because
its extent exceeds the formal 0.63~Mpc because of its
complex shape.

\onecolumngrid
 \renewcommand\baselinestretch{0.74}
\setcaptionmargin{0mm} \onelinecaptionstrue
\captionstyle{nonumber}
\medskip
\begin{longtable}{c|c|r|c|c|c|c|c|c}
\caption{\centerline{List of giants}} \\
 \hline
  $\alpha_{2000}$ & $\delta_{2000}$ & Angular   & Total   &  Flux density  & g-band       & z & Apparent  &
                  Object\\
                  &                 &  size,    & flux    &  of the central& magnitude    &   & size,     & type  \\
                  &                 & arcsec    &density, &  component,    &of the optical&   & Mpc       &       \\
                  &                 &           & mJy     &  mJy           &  component   &   &           &       \\
  \hline
(1)&    (2)& \multicolumn{1}{c|}{(3)}& (4)& (5)& (6)& (7)& (8)& (9)\\

\hline \multicolumn{9}{c}{NVSS}\\ [2pt]
\endfirsthead
\caption{\centerline{(Cont.)}}\\
\hline
  $\alpha_{2000}$ & $\delta_{2000}$ & Angular   & Total   &  Flux density  & $g$-band     & z & Apparent  & Object\\
                  &                 &  size,    & flux    &  of the central& magnitude    &   & size,     & type  \\
                  &                 & arcsec    &density, &  component,    &of the optical&   & Mpc       &       \\
                                    &           & mJy     &  mJy           & component    &   &           & 
                  \\

  \hline
(1)&    (2)& \multicolumn{1}{c|}{(3)}& (4)& (5)& (6)& (7)& (8)& (9)\\
\hline
\endhead
\hline
\endfoot
\hline
\endlastfoot
 %NVSS
 00 03 31.50&  $+03~51~11.0$&  1174.7&   470.6&    10.9&  17.11&  0.097&  2.08&  Gal\\
 00 17 47.80&  $-22~23~20.0$&   386.5&   485.6&     0.1&  17.40&  0.108&  0.75&  Gal\\
 00 20 32.90&  $-20~16~13.0$&   490.5&   965.8&   207.6&  18.80&  0.197&  1.58&  Gal\\
 00 37 19.10&  $+26~13~12.0$&   435.1&   156.0&   119.4&  17.43&  0.148&  1.12&  Gal\\
 00 43 59.80&  $+31~37~20.0$&   159.9&   129.3&    11.3&  17.85&  0.631&  1.09&  QSO\\
 01 20 38.60&  $-17~01~55.0$&   505.1&   310.4&    81.5&  17.22&  0.089&  0.83&  Gal\\
 01 43 56.00&  $+06~24~38.0$&   268.8&   134.9&    29.3&  17.99&  0.180&  0.81&  Gal\\
 01 55 46.30&  $-26~54~04.8$&   220.4&    22.7&     8.3&  17.27&  0.209&  0.75&  Gal\\
 02 49 36.80&  $-20~30~11.0$&   480.0&  1020.6&     0.1&  16.96&  0.087&  0.77&  Gal\\
 02 51 46.80&  $+15~50~13.0$&   131.8&   763.7&    30.1&  17.58&  0.489&  0.79&  QSO\\
 03 11 51.70&  $-31~30~01.0$&   106.3&  1052.0&   780.3&  20.51&  2.420&  0.88&  QSO\\
 03 13 32.90&  $-06~31~58.0$&   177.3&   164.8&     0.1&  19.50&  0.389&  0.93&  Gal\\
 04 22 21.00&  $+15~10~59.0$&   735.0&   168.8&     6.7&  18.10&  0.409&  3.98&  Gal\\
 07 46 33.70&  $+17~08~10.0$&   436.5&    81.2&    16.7&  19.21&  0.188&  1.36&  Gal\\
 07 53 41.30&  $+34~30~32.0$&   252.8&   111.7&    11.6&  23.59&  0.548&  1.61&  Gal\\
 08 24 01.00&  $+24~47~36.0$&   301.2&    39.0&     0.1&  19.14&  0.224&  1.08&  Gal\\
 08 32 34.10&  $+04~24~36.0$&   446.8&   191.9&   112.5&  15.95&  0.106&  0.86&  Gal\\
 08 41 45.90&  $-25~38~18.0$&   619.9&    76.8&    22.0&  14.68&  0.083&  0.95&  Gal\\
 08 45 25.51&  $+52~29~15.9$&   211.7&    41.9&    20.1&  20.74&  0.403&  1.14&  Gal\\
 08 48 36.10&  $+05~55~23.0$&   157.4&  1319.5&    87.0&  19.06&  0.798&  1.18&  Gal\\
 08 57 01.76&  $+01~31~30.9$&   276.3&    96.3&     8.1&  20.25&  0.273&  1.15&  Gal\\
 09 11 54.60&  $+08~12~31.0$&   203.6&   347.8&    86.3&  18.64&  0.243&  0.77&  Gal\\
 09 12 36.60&  $+68~34~25.0$&   290.6&   425.4&   396.6&  18.69&  1.080&  2.38&  Gal\\
 09 14 19.50&  $+10~06~41.0$&   366.1&   469.8&   348.7&  19.80&  0.311&  1.65&  Gal\\
 09 32 38.30&  $+16~11~57.0$&   236.7&   748.9&     1.2&  18.43&  0.191&  0.75&  Gal\\
 09 40 03.80&  $+51~04~22.0$&   240.7&   254.7&    16.3&  18.87&  0.207&  0.81&  Gal\\
 09 54 19.19&  $+27~15~59.9$&   186.7&   150.5&    11.0&  20.65&  0.471&  1.10&  Gal\\
 10 14 43.92&  $-01~46~12.0$&   240.8&   225.0&    31.7&  18.02&  0.198&  0.78&  Gal\\
 10 17 18.70&  $+39~31~21.0$&   125.9&   745.4&    13.6&  20.20&  0.530&  0.79&  Gal\\
 10 36 23.00&  $+38~31~31.1$&   185.0&   128.8&     2.7&  20.36&  0.408&  1.00&  Gal\\
 10 48 04.80&  $+74~19~40.0$&   643.5&    69.2&     0.1&  17.29&  0.121&  1.39&  Gal\\
 11 00 38.30&  $-02~34~37.0$&   227.7&    71.8&    41.0&  21.39&  0.399&  1.21&  Gal\\
 11 04 46.99&  $+21~03~17.7$&   237.0&   108.1&     7.7&  18.53&  1.153&  1.96&  QSO\\
 11 18 59.17&  $+27~54~49.5$&   180.9&   523.0&     0.1&  20.48&  0.317&  0.83&  Gal\\
 11 47 20.70&  $-12~53~10.0$&   125.0&   364.4&     0.1&  16.93&  0.497&  0.76&  Gal\\
 11 48 55.90&  $-04~04~10.0$&   190.3&   613.9&   175.3&  17.17&  0.340&  0.91&  QSO\\
 11 53 18.00&  $+03~38~05.0$&   170.9&    52.7&     5.8&  20.71&  0.328&  0.80&  Gal\\
 11 56 54.70&  $+26~32~32.0$&   271.6&   168.9&   124.2&  18.09&  0.156&  0.72&  Gal\\
 11 59 26.20&  $+21~06~56.0$&   156.9&   270.2&    21.2&  17.27&  0.349&  0.77&  QSO\\
 12 02 40.30&  $-12~51~40.0$&   198.3&   131.8&    93.4&  16.56&  0.553&  1.27&  QSO\\
 12 13 57.10&  $+08~32~02.0$&   158.2&    45.6&    13.7&  17.87&  0.811&  1.20&  QSO\\
 12 48 13.90&  $+36~24~24.0$&   251.0&    90.5&     2.9&  18.13&  0.207&  0.84&  Gal\\
 12 55 50.13&  $+58~18~42.0$&   216.4&    34.7&    12.8&  20.52&  0.361&  1.08&  Gal\\
 13 13 57.70&  $+64~25~55.0$&   248.6&    78.0&     0.1&  18.99&  0.221&  0.88&  Gal\\
 13 14 03.40&  $-33~03~56.0$&   185.4&   233.6&    95.1&  19.43&  0.484&  1.11&  QSO\\
 13 14 43.80&  $+27~37~41.0$&   236.9&    30.1&     5.9&  20.93&  0.418&  1.30&  Gal\\
 13 20 41.10&  $+45~51~23.0$&   306.7&    84.1&    12.5&  17.93&  0.178&  0.91&  Gal\\
 13 28 34.10&  $-01~29~18.0$&   308.4&   354.4&    15.6&  17.91&  0.151&  0.80&  Gal\\
 13 28 34.40&  $-03~07~45.0$&   804.7&   207.5&    12.6&  17.49&  0.085&  1.27&  Gal\\
 13 45 57.50&  $+54~03~17.0$&   259.4&   346.0&     9.3&  18.35&  0.163&  0.72&  Gal\\
 13 50 36.10&  $-16~34~50.0$&  1153.3&   300.3&   109.2&  16.60&  0.098&  2.06&  QSO\\
 13 53 35.92&  $+26~31~47.5$&   158.0&   239.6&    55.3&  16.87&  0.310&  0.71&  QSO\\
 14 06 26.00&  $+45~09~05.0$&   263.6&    44.2&     2.2&  20.85&  0.400&  1.41&  Gal\\
 14 22 49.20&  $-27~27~56.0$&   102.8&  2418.2&  2268.8&  17.60&  0.985&  0.82& QSO\\
 14 30 45.66&  $+14~50~38.0$&   259.5&    72.0&     5.2&  16.41&  0.377&  1.33&  QSO\\
 14 31 51.10&  $+10~29~59.0$&   270.4&    61.8&    18.5&  18.11&  0.166&  0.76&  Gal\\
 14 32 44.90&  $+30~14~35.0$&   183.9&    75.4&    50.6&  18.81&  0.355&  0.91&  QSO\\
 14 41 24.00&  $-34~56~46.0$&   156.7&   419.6&   395.1&  17.43&  1.159&  1.30&  QSO\\
 14 47 10.20&  $+22~10~07.6$&   197.7&    49.3&     0.1&  18.99&  0.249&  0.77&  Gal\\
 14 49 54.80&  $+14~06~11.0$&   298.1&   133.1&     7.7&  21.98&  0.251&  1.16&  Gal\\
 14 51 06.40&  $+53~33~54.0$&   187.9&    42.5&     2.6&  19.07&  0.432&  1.05&  Gal\\
 15 13 44.90&  $-10~12~00.0$&   228.5&  1062.8&   874.5&  18.40&  1.513&  1.95&  QSO\\
 15 18 30.90&  $+48~32~14.0$&   108.0&    66.4&     5.1&  18.35&  0.576&  0.71&  QSO\\
 15 24 44.60&  $+19~59~57.0$&   247.3&    25.8&     0.1&  20.39&  0.345&  1.20&  Gal\\
 15 29 18.00&  $+32~48~42.0$&   298.1&    65.7&    32.4&  17.75&  1.650&  2.55&  QSO\\
 15 29 50.80&  $+02~25~15.0$&   163.3&    88.1&     7.3&  20.63&  0.339&  0.78&  Gal\\
 15 40 56.80&  $-01~27~10.0$&   275.1&   214.9&    10.3&  17.78&  0.149&  0.71&  Gal\\
 15 48 16.70&  $+69~49~35.0$&   204.2&   272.7&    33.1&  16.65&  0.375&  1.05&  QSO\\
 15 52 06.67&  $+22~47~39.5$&   667.3&   545.5&     5.7&  17.42&  0.116&  1.39&  Gal\\
 15 52 22.70&  $+22~33~12.0$&   672.9&   265.5&     7.3&  15.72&  0.068&  0.86&  Gal\\
 16 03 34.10&  $+36~59~52.0$&   155.9&    45.8&     9.5&  18.40&  0.967&  1.24&  QSO\\
 16 09 53.40&  $+43~34~11.0$&   123.9&   337.1&     9.5&  17.61&  0.760&  0.91&  QSO\\
 16 22 06.00&  $+24~49~16.0$&   334.2&    58.7&     5.0&  18.41&  0.148&  0.86&  Gal\\
 16 23 42.40&  $+25~21~47.0$&   283.0&   226.3&    68.1&  21.50&  0.364&  1.42&  Gal\\
 16 49 28.90&  $+30~46~52.0$&   108.4&   110.6&    39.7&  18.62&  1.125&  0.89&  QSO\\
 17 52 46.00&  $+17~34~20.0$&   139.0&   370.6&   341.3&  18.10&  0.504&  0.85&  QSO\\
 19 21 14.00&  $+48~06~19.0$&   519.4&  1097.9&     0.1&  12.27&  0.102&  0.96&  Gal\\
 20 57 20.40&  $+00~12~07.0$&   267.7&   406.4&    74.9&  17.85&  0.135&  0.63&  Gal\\
 21 45 04.50&  $-06~59~08.0$&   341.8&   160.0&     9.7&  19.60&  0.315&  1.56&  Gal\\
 22 25 13.60&  $-16~19~04.0$&   626.2&   547.8&     0.1&  16.05&  0.103&  1.17&  Gal\\
 22 30 40.30&  $-39~42~52.0$&   191.3&   644.3&   320.1&  17.38&  0.318&  0.88&  QSO\\
 22 34 58.80&  $-02~24~18.0$&   192.1&    73.6&    24.8&  18.55&  0.550&  1.23&  QSO\\
 22 41 01.99&  $+27~32~59.0$&   269.4&   258.3&    51.5&  17.43&  0.493&  1.63&  QSO\\
 22 53 36.00&  $-34~55~31.0$&   257.8&   280.0&    50.0&  17.83&  0.212&  0.88&  Gal\\
 23 16 00.60&  $-28~23~53.0$&   437.9&   319.2&     5.6&  18.18&  0.229&  1.59&  Gal\\
 23 21 54.80&  $-24~25~52.0$&   179.4&    43.5&    10.1&  18.21&  0.279&  0.76&  Gal\\
 23 33 12.40&  $+00~56 58.5$&   890.4&    87.3&     0.1&  16.43&  0.087&  1.43&  Gal\\
 23 33 55.24&  $-23~43~40.7$&  1189.9&  1118.5&   782.1&  13.82&  0.048&  1.10&  Gal\\
 23 45 33.10&  $-04~48~54.6$&  1003.6&   161.6&     0.1&  15.72&  0.075&  1.41&  Gal\\
 23 55 24.70&  $+26~24~14.0$&   195.9&   316.0&    14.8&  19.44&  0.240&  0.74&  Gal\\[3pt]
 \multicolumn{9}{c}{FIRST}\\[3pt]
 08 38 13.13&  $+13~58~10.7$&   127.9&    59.9&    3.7&   18.72&  2.024&  1.08&  QSO\\
 09 42 16.50&  $+12~45~03.6$&    93.0&    46.6&   12.0&   19.09&  1.432&  0.79&  QSO\\
 10 17 54.86&  $+47~05~29.0$&   116.1&    25.2&    7.8&   18.69&  0.668&  0.81&  QSO\\
 10 20 03.00&  $-02~47~18.0$&    95.0&    88.5&    6.7&   20.53&  1.447&  0.81&  QSO\\
 10 38 22.47&  $+13~46~57.0$&   101.0&    72.2&   65.6&   17.66&  0.947&  0.80&  QSO\\
 10 56 37.96&  $+27~43~43.8$&    94.2&   316.5&   15.4&   20.49&  0.998&  0.76&  QSO\\
 11 09 35.40&  $+51~04~02.6$&    92.4&    30.2&    3.1&   19.48&  1.179&  0.77&  QSO\\
 11 12 15.45&  $+11~29~19.3$&   108.4&    74.0&    7.7&   19.15&  1.132&  0.90&  QSO\\
 12 08 22.00&  $+22~19~58.2$&   108.8&   135.0&    7.9&   18.45&  0.745&  0.80&  QSO\\
 12 14 31.15&  $+18~28~15.1$&    89.8&   100.6&   17.7&   20.58&  1.590&  0.77&  QSO\\
 12 57 10.80&  $+40~54~29.2$&   119.5&    29.5&    2.9&   19.30&  1.067&  0.98&  QSO\\
 13 33 07.00&  $+04~50~48.5$&   129.5&    48.0&    6.3&   18.50&  1.405&  1.10&  QSO\\
 13 56 00.04&  $+19~04~20.8$&    90.8&    36.0&   12.5&   18.62&  2.224&  0.76&  QSO\\
 14 15 54.37&  $+49~09~21.2$&    89.8&    25.6&   12.0&   19.60&  1.371&  0.76&  QSO\\
 14 37 47.74&  $+07~48~56.2$&   139.5&    56.0&    3.1&   18.78&  1.472&  1.19&  QSO\\
 14 39 32.68&  $+45~50~28.3$&   127.1&    21.7&    3.2&   19.24&  1.836&  1.08&  QSO\\
 14 46 26.80&  $+41~33~18.0$&   102.8&   496.2&    3.0&   20.44&  0.675&  0.72&  Gal\\
 14 50 38.83&  $+45~49~54.6$&    93.0&    93.8&    9.2&   19.46&  1.622&  0.80&  QSO\\
 14 51 03.23&  $+11~41~08.6$&    88.3&    37.6&   13.2&   18.44&  1.067&  0.72&  QSO\\
 14 53 08.00&  $+22~17~07.7$&   115.9&   103.4&   15.8&   19.74&  0.785&  0.87&  QSO\\
 15 07 39.50&  $+11~04~03.7$&   140.8&   226.4&    8.8&   18.02&  0.475&  0.83&  QSO\\
 15 33 13.22&  $+06~58~01.6$&    95.6&    45.6&   33.6&   18.23&  1.160&  0.79&  QSO\\
 15 49 33.46&  $+00~47~32.6$&    84.2&    20.2&   12.0&   19.52&  1.253&  0.71&  QSO\\
 16 01 51.57&  $+17~54~10.2$&   104.6&   166.6&    5.6&   18.27&  0.660&  0.73&  QSO\\
 16 23 46.42&  $+27~35~13.6$&    93.8&    63.6&   57.5&   18.82&  1.397&  0.80&  QSO\\
 23 56 06.32&  $-01~31~51.2$&    98.2&   190.5&    6.1&   22.23&  1.028&  0.80&  Gal\\
 \end{longtable}
 \renewcommand\baselinestretch{1.0}
\twocolumngrid

\begin{figure}[]
\captionstyle{normal}\setcaptionmargin{5mm} \onelinecaptionstrue
\includegraphics[scale=0.3,angle=90]{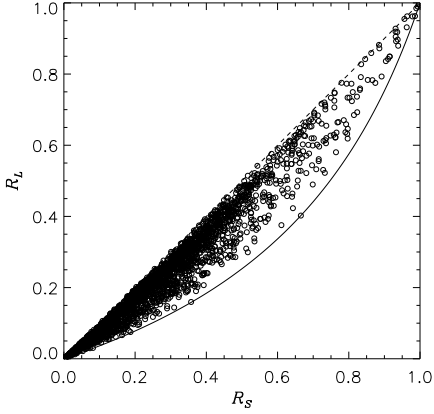}
\caption{Relation between   $R_L$ and $R_S$ for the radio sources of
the first list. The dashed line shows the $R_L$~=~$R_S$ relation. The
lower boundary of the plot (the solid line) was computed by
formula~(1) for $z=3.7$ (maximum redshift in this sample).}
 \label{fig_1:Amirkhanyan_n_en}
\end{figure}

\section{DETERMINATION OF THE PARAMETERS OF THE RADIATION PATTERN}

\subsection{The First Sample, $D < 0.7$~Mpc}

For all radio sources we determine parameter $R_L$ by formula~(1).
Figure~1 shows the relation between $R_L$ and $R_S$. The dashed line
shows the identity relation  $R_L$~=~$R_S$. The lower boundary of the
plot (the solid line) was computed by formula~(1) for \mbox {$z=3.7$}
 (the maximum redshift in the sample). We adopted the same
 spectral indices for all radio sources because these data are
 unavailable for most of the radio sources. We set $\alpha_E=-0.8$
 and $\alpha_C=-0.1$ for extended and compact components, respectively.
The eventual selection by redshift appears to be small because the
average $z$ of radio sources weakly depends on $R_L$ (Fig.~2).

 \setcaptionmargin{5mm}
\onelinecaptionstrue \captionstyle{normal}
\begin{figure}[]
\includegraphics[scale=0.3,angle=90]{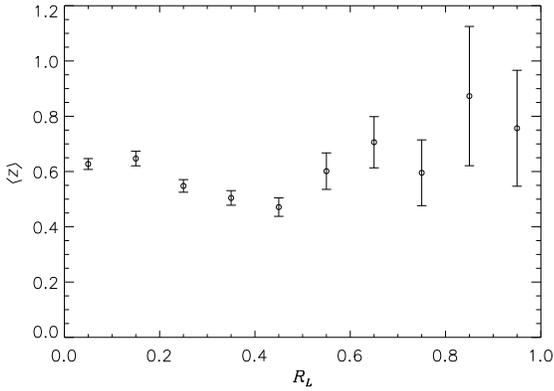}
\caption{Dependence of $z$ on $R_L$ for radio sources with apparent
sizes $D < 0.7$~Mpc. We subdivided the entire range of  $R_L$ values
into ten equal bins. In each bin we computed the mean redshift by
averaging the $z$ values of the objects and the corresponding
standard deviation.}
 \label{fig_2:Amirkhanyan_n_en}
\end{figure}

\setcaptionmargin{5mm} \onelinecaptionsfalse \captionstyle{normal}
\begin{figure*}[]
\includegraphics[scale=0.3,angle=90]{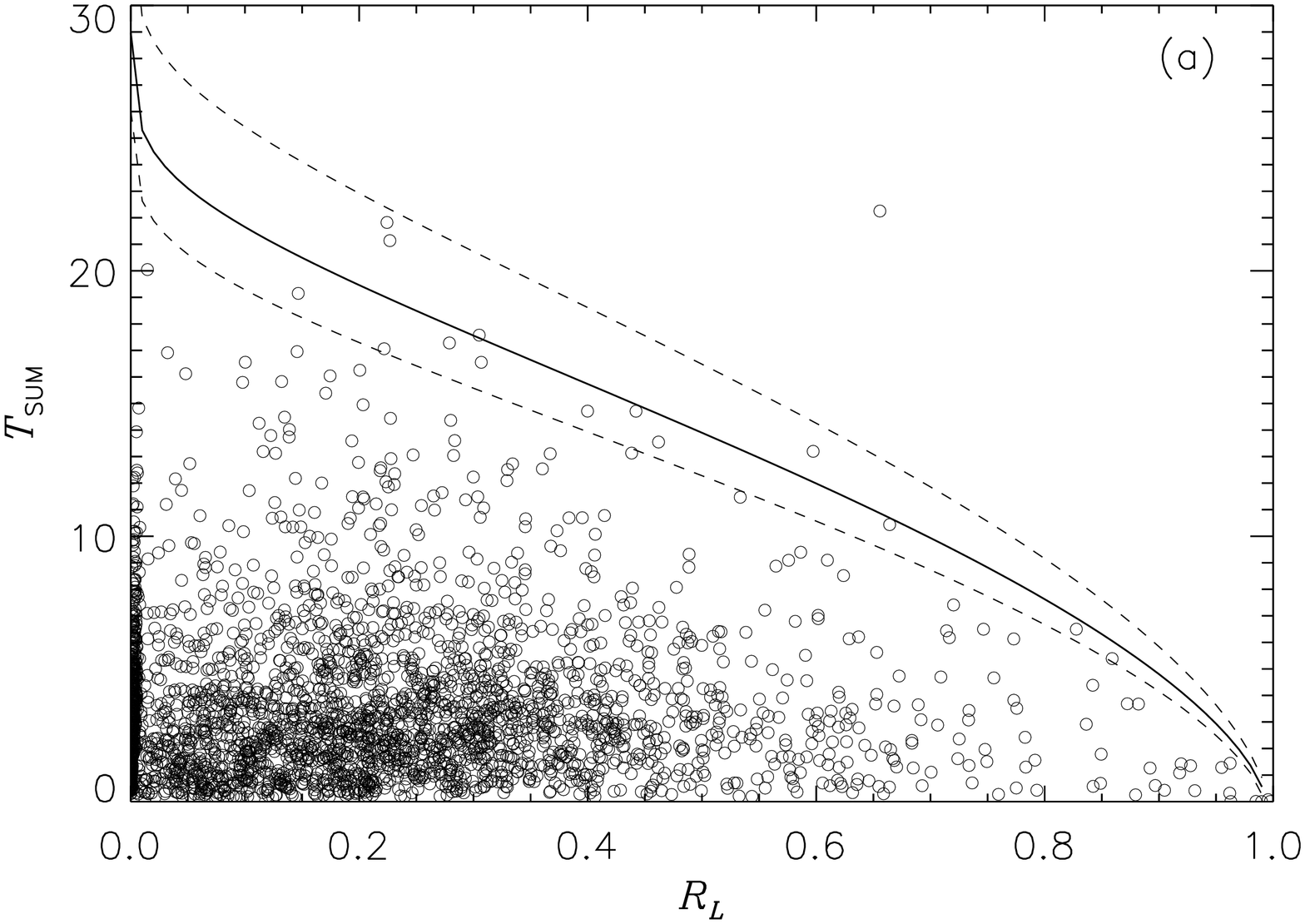} 
\includegraphics[scale=0.3,angle=90]{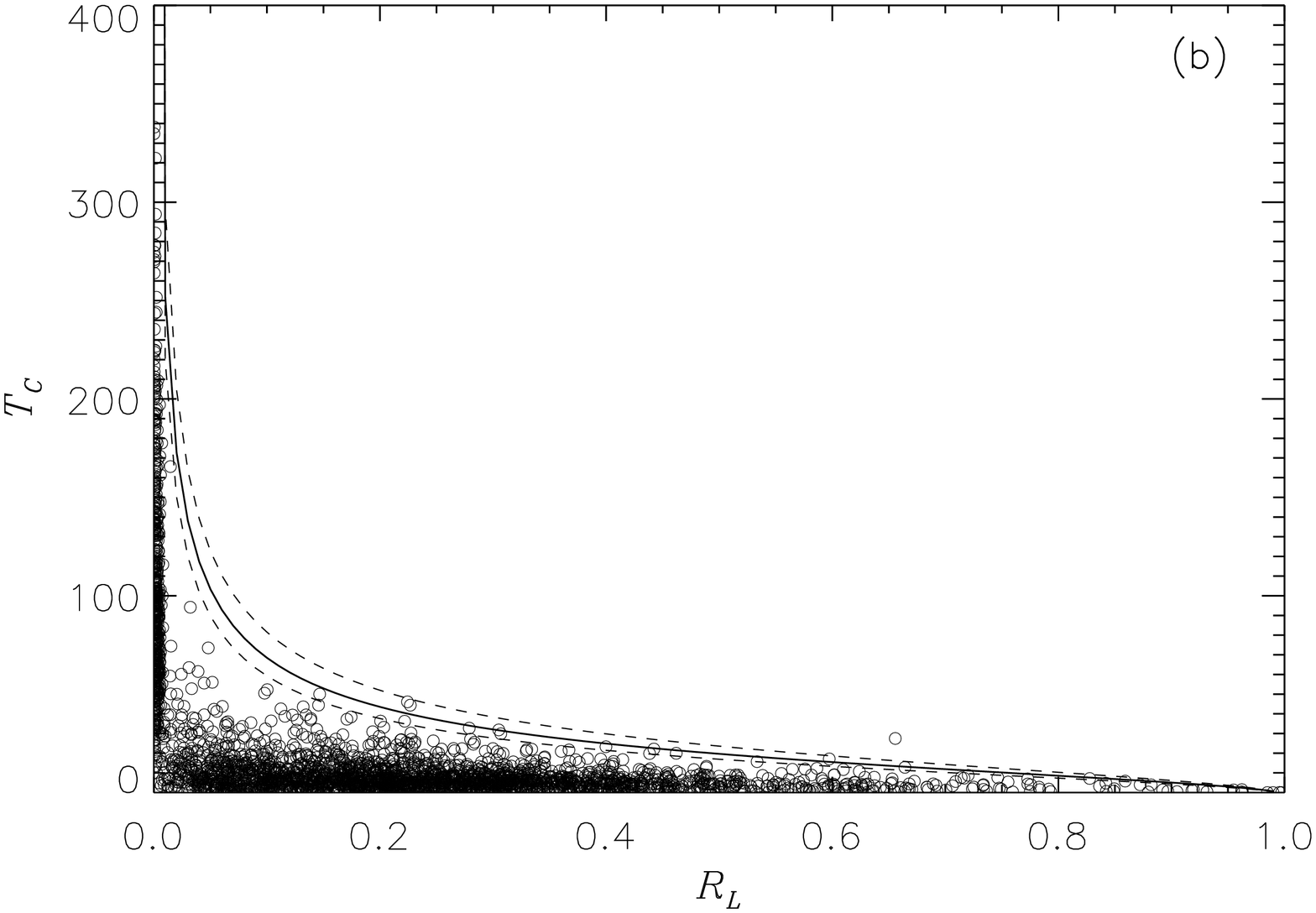}\\ 
\includegraphics[scale=0.3,angle=90]{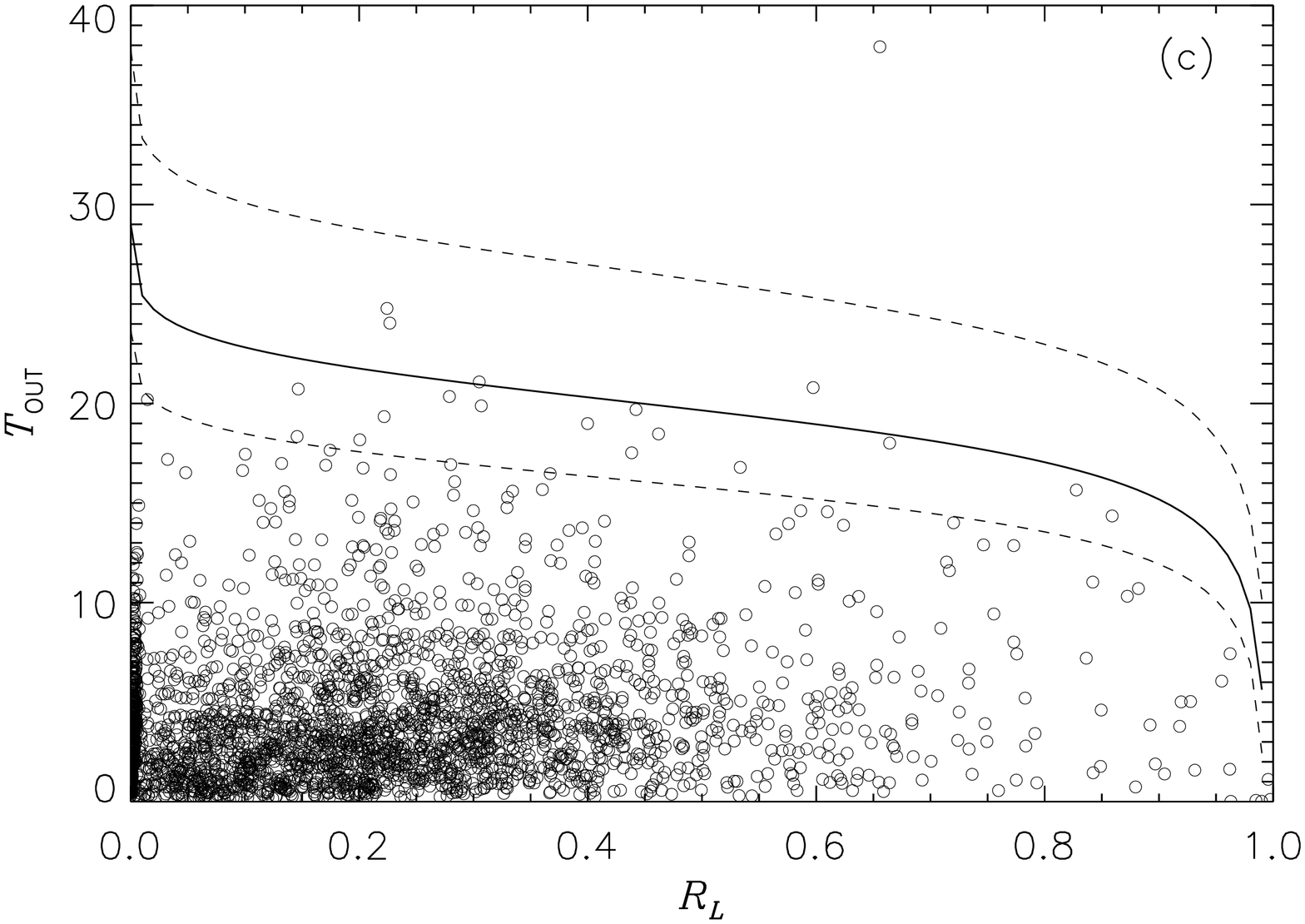} 
\includegraphics[scale=0.3,angle=90]{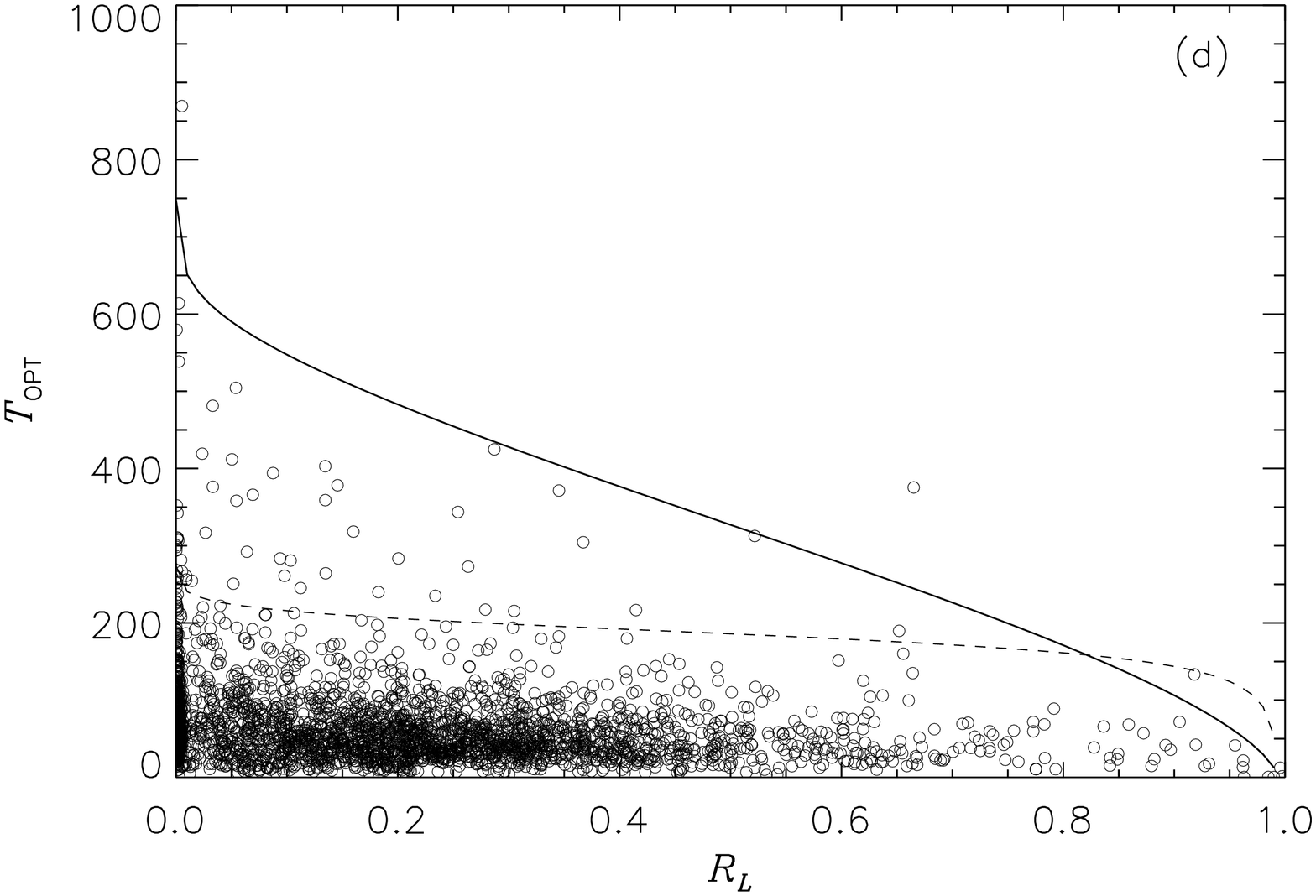}\\ 
\caption{ Radio sources with apparent sizes smaller than 0.7~Mpc.
Dependence on $R_L$: (a)  $T_{\rm SUM}$, (b) $T_C$,  (c) $T_E$, (d)
$T_{\rm opt}$. The solid line in Figs.~(a), (b), and (c) shows the
upper limit computed by formulas~(6)--(8) for the optimum parameters
of the RP of the radio source: \mbox {$a=0.007$}, \mbox {$n=15$}. The
upper limit in Fig.~(d) is computed by formula~(9) for the parameters
of the RP of the optical component \mbox {$a_{\rm opt}=0.005$}, \mbox
{$n_{\rm opt}=15$}. The dashed line shows the boundary of spherical
RP.}
\end{figure*}

We now determine parameter $T$ for different radio-source components
by substituting into formula~(3) the total flux densities $S_{\rm
SUM}$, flux densities $S_C$ of the central fraction, or flux
densities $S_E$ of external components.  \mbox{Figs.~3a,~3b, and~3c}
show the plots  $T_{\rm SUM}$, $T_C$, and $T_E$ as functions of $R_L$
for objects of the first  sample.

The distribution functions of the true sizes and true luminosities of
extended radio sources do not depend on their spatial orientation and
therefore the upper boundary of parameter $T$ as a function of $R_L$
is determined by the RP of radio sources (formulas~\mbox{(6)--(8)}).
When computing the upper limit (the solid line in Figs.~3a--3c) we
set the maximum true sized of radio sources equal to \mbox {$D_0 =
0.7$}~Mpc.

We then varied $a$, $n$, and $I_0$ to determine the parameter values
that allow achieving the best agreement between the computed and
visible boundaries simultaneously in three plots: \mbox {$a=0.007$},
\mbox {$n=15$},
 \mbox {$I_0=10^{25}$~W\,Hz$^{-1}$\,sr$^{-1}$}.

\setcaptionmargin{5mm} \onelinecaptionsfalse \captionstyle{normal}
\begin{figure*}[]
\includegraphics[scale=0.3,angle=90]{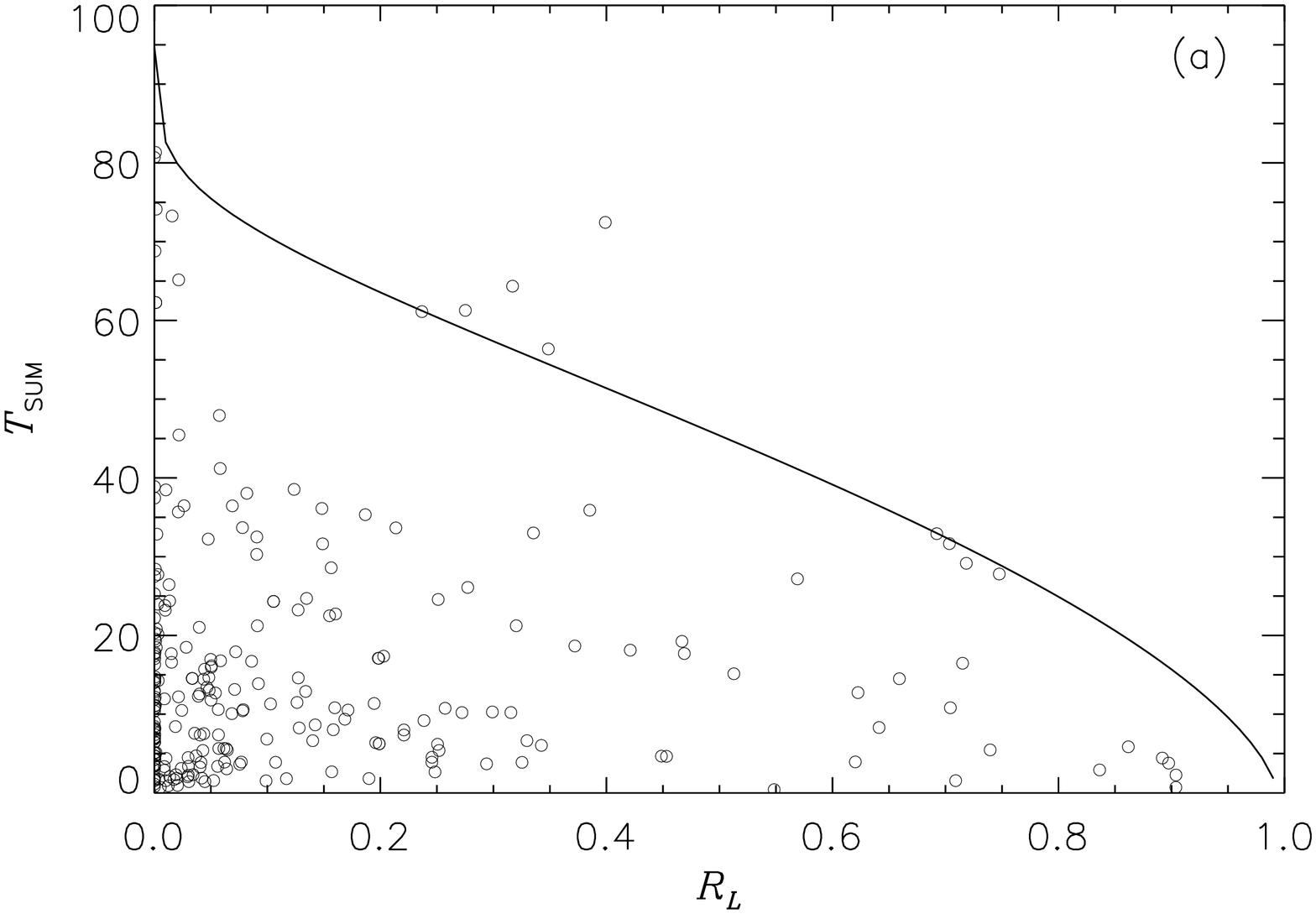} 
\includegraphics[scale=0.3,angle=90]{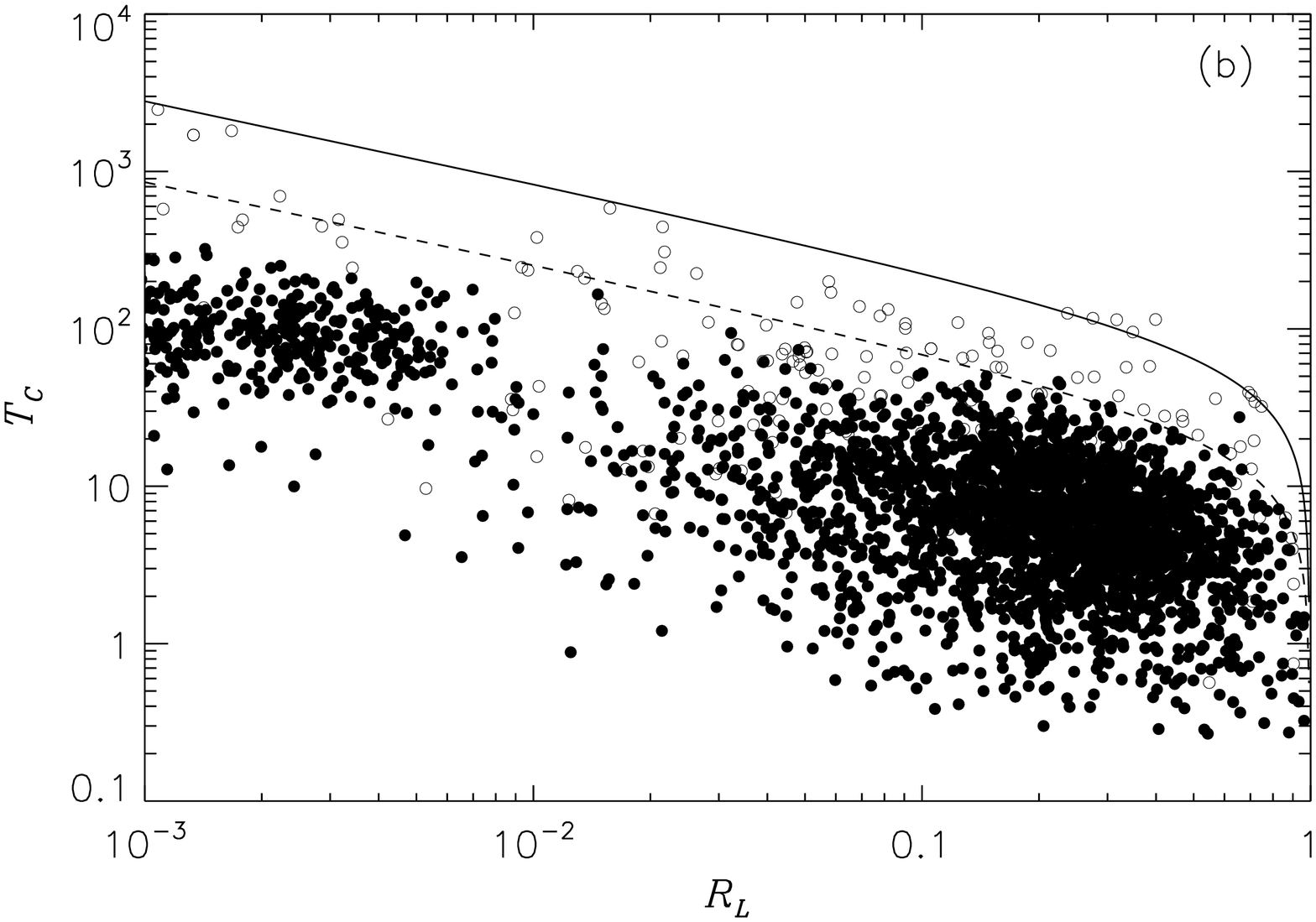}\\ 
\includegraphics[scale=0.3,angle=90]{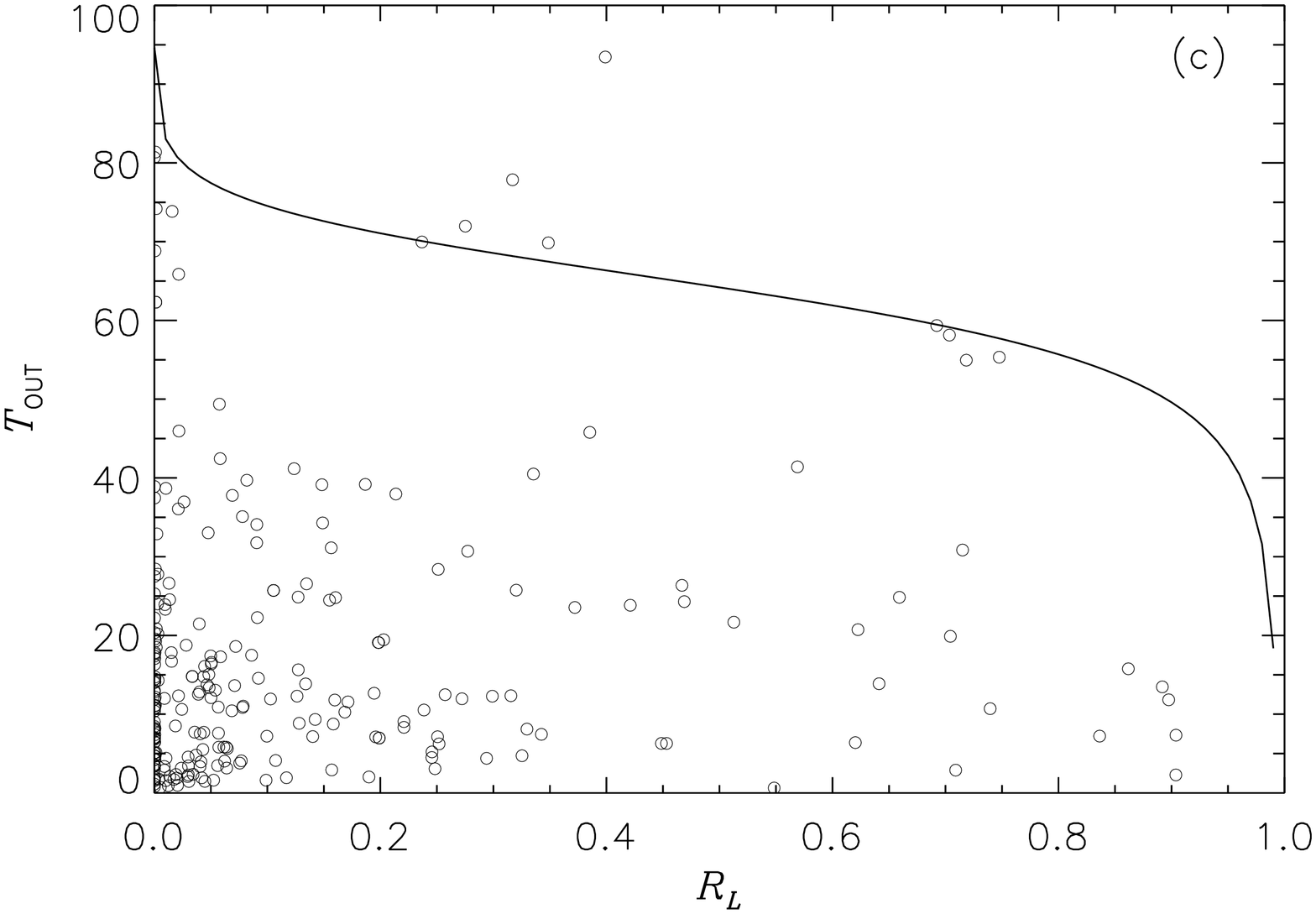} 
\includegraphics[scale=0.3,angle=90]{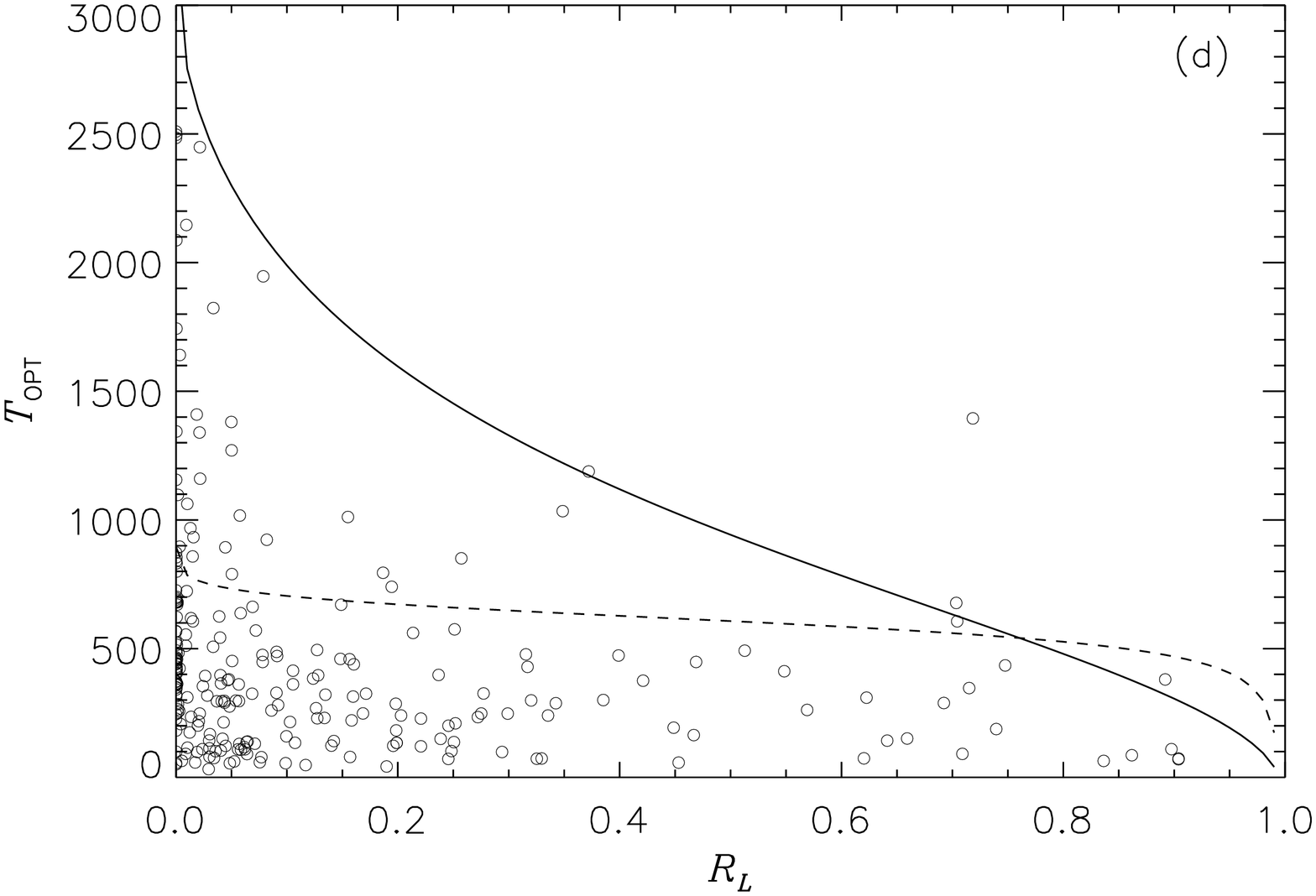}\\ 
\caption{Radio sources with apparent sizes greater than 0.7~Mpc.
Dependence on  $R_L$: (a) $T_{\rm SUM}$, (b) $T_C$, (c) $T_E$, (d)
and $T_{\rm opt}$. The solid line in Figs.~a, b, c shows the upper
limit computed by formulas~(6)--(8) for parameters of the
radio-source RP \mbox {$a=0.007$}, \mbox {$n=15$}. The dashed line in
(b) shows, for comparison, the boundary of objects of the first list
(the filled circles). The upper boundary in (d) was computed by
formula~(9) for parameters of the RP of the optical component \mbox
{$a_{\rm opt}=0.003$}, \mbox {$n_{\rm opt}=13$}. The dashed line
shows the boundary of the spherical RP.}
\end{figure*}

The normalization of  RP~(4) differs from traditional $4\pi$:
\begin{equation}
      N=4\pi\left(a+ \frac {1-a} {2n+1}\right),
\end{equation}
and is equal to 0.49 for the inferred RP parameters. Hence the
minima, luminosity of the radio sources of the sample is  \mbox
{$N\times I_0 =4.9\times 10^{24}$~W\,Hz$^{-1}$}.

The dashed line in Figs.~3a and~3c demonstrate the sensitivity of the
computed upper boundaries for parameter values from \mbox {$a=0.005$}
to \mbox {$a=0.009$}. For the same purpose we show by the
 dashed line in Fig.~3b the boundary for $n=12$ and $n=18$. Figures~3a, 3b and~3c
 clearly demonstrate that the RP of the extended source is evidently non-spherical.
 Its main lobe whose maximum coincides with the major axis of the radio source
 has the width of about $24\degr$.
The distribution of the radiation of outer components is close to spherical and its
level is 140 times lower than the intensity at the maximum and the luminosity of
the compact component is a factor of 5.48 higher than that of outer
components.

Before computing parameter $T$ of the optical component we must
decide how to compute the \mbox{$K$-correction} for the AGN. The
\mbox {$K$-correction} is usually viewed as a magnitude correction.
Here we define it as the $g$-band flux density correction for the
optical component of the radio source. To determine the \mbox
{$K$-correction}, we used the most complete and homogeneous
photometric measurements in the $u$-, $g$-, $r$-, $i$-, and
$z$-filters from the \mbox
{SDSS-survey}\footnote{http://skyserver.sdss.org/dr12/en/tools/crossid/\\crossid.aspx.}.
For  540 objects with redshifts greater than 1 we supplemented these
data with infrared photometry from the  2MASS (\mbox{$J$-},
\mbox{$H$-}, and \mbox {$K$-filters})~\cite{27:Amirkhanyan_n_en}
 and  WISE All-sky (\mbox {$W1$-filter}) survey~\cite{28:Amirkhanyan_n_en}.
We then constructed the spectrum of the optical component based on
the entire set of photometric data and determined the \mbox
{$K$-correction}.

To compute   $T_{\rm opt}$ we rewrite equation~(3) taking into
account the inferred \mbox {$K$-correction} :
\begin{equation}
T_{\rm opt}=\frac{\Theta}{\sqrt{\mathstrut S_g K_{\rm
COR}}}(1+z)^{-2}.
\end{equation}
Here $S_g$ is the observed flux density of the object in the $g$-band
filter. We show the results of computations in Fig.~3d.
Unexpectedly, the plot showed that the RP of the optical component of the
radio source is far from spherical. The solid line
shows the optimum upper boundary for \mbox {$a_{\rm
opt}=0.005$}, \mbox {$n_{\rm opt}=15$}, \mbox {$I_{\rm
0,\,opt}=2\times10^{22}$}~W\,Hz$^{-1}$\,sr$^{-1}$ . The dashed
line shows where the boundary should be in the case of spherical RP in the
optical range with intensity $I_{\rm 0,\,opt}$.

 \subsection{Second Sample, $D > 0.7$~Mpc}

The computations performed for objects of the first sample were
repeated for the giant radio sources of the second sample.
Figures~4a,~4b,~4c, and~4d show the results of computations for
$T_{\rm SUM}$, $T_C$, $T_E$, and $T_{\rm opt}$ (the open circles).
The boundary of the distribution was computed for  \mbox {$a=0.007$},
\mbox {$n=15$},
\mbox{$I_0=1.5\times10^{25}$~W\,Hz$^{-1}$\,sr$^{-1}$},  \mbox {$D_0 =
2.8$}~Mpc. It is evident that because of the limited statistics the
agreement with experimental data is not so good as in  the case of
the first sample. It is, however, evident that the RP of giant radio
sources is also non-spherical and its parameters are close to those
inferred for the first list. To demonstrate the latter statement, we
added to Fig.~4b objects from Fig.~3b (the filled circles). To cover
the wide range of  $T_C$ and $R_L$ values, the figure is plotted in
logarithmic scale. The relative shift of the boundaries of the first
and second samples along the vertical axis (the dashed and solid
lines, respectively) is primarily determined by the change of the
true maximum size of radio sources and not the parameters of the RP.
We could find photometry in the SDSS catalog for computing $T_{\rm
opt}$ (Fig.~4d) only for 146 giant radio sources. We compiled the
$BVRIJHK$ photometry from VizieR
database\footnote{http://vizier.u-strasbg.fr/viz-bin/VizieR.}. This
allowed us to determine the $g$-band flux density and the
corresponding $K$-correction reduced to this band for 254 objects.
The boundary was computed for the following parameter values: \mbox
{$a_{\rm opt}=0.003$},    \mbox {$n_{\rm opt}=13$}, \mbox {$I_{\rm
0,\,opt}=3\times10^{22}$}~W\,Hz$^{-1}$\,sr$^{-1}$, \mbox
{$D_0=2.8$~Mpc}\linebreak(Fig.~4d).

The dashed line in the same figure shows where the boundary should be
if the RP in the optical is spherical and \mbox{$I_{\rm
0,\,opt}=3\times10^{22}$}~W\,Hz$^{-1}$\,sr$^{-1}$. It follows from
these parameters that the FWHM of the main lobe of the optical RP is
$26\fdg4$.

\section{CONCLUSIONS}

 It can be concluded from the above that:
 \begin{list}{}{
\setlength\leftmargin{2mm} \setlength\topsep{2mm}
\setlength\parsep{0mm} \setlength\itemsep{2mm} }
 \item (1) The ratio of the radiation of the compact component  
 to total radiation  of the extendet radio source is indeed 
 connects with its spatial orientation.
 \item (2) The form of the RP does not depend on the size or luminosity
 of the radio source. This may also be true for objects with luminosities
 higher than
\mbox{$L=4.9\times10^{24}$~W\,Hz$^{-1}$}.
 \item (3) The central component emits within a narrow beam whose width at  1.4~GHz
 is of about $24\degr$. This value corresponds to  \mbox{$\gamma=2.33$}.
 \item (4) The RP of the extended component is close to spherical, its level
 is about  \mbox {0.005--0.01} of the intensity at the maximum, and its
 luminosity is equal to  0.13--0.24 that of the entire radio source.
 \item (5) The RP of the optical component of the radio source is also
 non-spherical, its radiation is concentrated within a beam of width
\mbox{$24$--$27\degr$}, and the level of the spherical component
is of about \mbox{0.003--0.005} of the intensity at the maximum.
\end{list}

\begin{acknowledgements}
This research has made use of the VizieR catalogue access tool, CDS,
 Strasbourg, France.
\end{acknowledgements}


\begin{thebibliography}{99}
\bibitem{1:Amirkhanyan_n_en}
I.~S.~{Shklovskii}, \azh \textbf{42}, 30 (1965).

\bibitem{2:Amirkhanyan_n_en}
H.~{van der Laan}, \nat\ \textbf{211}, 1131 (1966).

\bibitem{3:Amirkhanyan_n_en}
M.~J.~{Rees}, \nat\ \textbf{211}, 468 (1966).

\bibitem{4:Amirkhanyan_n_en}
L.~M.~{Ozernoy} and V.~N.~{Sazonov}, \apss\ \textbf{3}, 365
(1969).

\bibitem{5:Amirkhanyan_n_en}
V.~N.~{Kuril'chik}, \azh \textbf{48}, 684 (1971).

\bibitem{6:Amirkhanyan_n_en}
Y.~A.~{Kovalev} and V.~P.~{Mikhailutsa}, \sovast \textbf{24}, 400
(1980).

\bibitem{7:Amirkhanyan_n_en}
P.~A.~G.~{Scheuer} and A.~C.~S.~{Readhead}, \nat\ \textbf{277},
182 (1979).

\bibitem{8:Amirkhanyan_n_en}
Y.~Y.~{Kovalev}, \azh \textbf{71}, 846 (1994).

\bibitem{9:Amirkhanyan_n_en}
S.~{Horiuchi}, D.~L.~{Meier}, R.~A.~{Preston}, and S.~J.~{Tingay},
\pasj  \textbf{58}, 211 (2006).

\bibitem{10:Amirkhanyan_n_en}
M.~J.~L.~{Orr} and I.~W.~A.~{Browne}, \mnras\ \textbf{200}, 1067
(1982).

\bibitem{11:Amirkhanyan_n_en}
V.~R. {Amirkhanyan}, \azh \textbf{70}, 16 (1993).

\bibitem{12:Amirkhanyan_n_en}
V.~R.~{Amirkhanyan}, \ab \textbf{69}, 383 (2014).

\bibitem{13:Amirkhanyan_n_en}
J.~J.~{Condon}, D.~T.~{Frayer}, and J.~J.~{Broderick}, \aj\
\textbf{101}, 362  (1991).

\bibitem{14:Amirkhanyan_n_en}
I.~W.~A.~{Browne} and R.~A.~{Battye}, ASP Conf. Ser. {\bf 427},
365 (2010).

\bibitem{15:Amirkhanyan_n_en}
V.~R.~{Amirkhanyan}, \ab \textbf{64}, 333 (2009).

\bibitem{16:Amirkhanyan_n_en}
K.~{Nilsson}, M.~J.~{Valtonen}, J.~{Kotilainen}, and
T.~{Jaakkola}, \apj \textbf{413}, 453 (1993).

17. L. Lara, W. D. Cotton, L. Feretti, et al., Astron. Astrophys., {\bf 370}, 409 (2001).\\

\bibitem{18:Amirkhanyan_n_en}
L.~{Lara}, I.~{M{\'a}rquez}, W.~D.~{Cotton},et~al.,\aaa \textbf{378},
826(2001).

\bibitem{19:Amirkhanyan_n_en}
C.~H.~{Ishwara-Chandra} and D.~J.~{Saikia}, \mnras\ \textbf{309},
100 (1999).

\bibitem{20:Amirkhanyan_n_en}
A.~P.~{Schoenmakers}, A.~G.~{de Bruyn}, H.~J.~A.~{R{\"o}ttgering},
and H.~{van  der Laan}, \aaa\ \textbf{374}, 861 (2001).

\bibitem{21:Amirkhanyan_n_en}
J.~{Machalski}, M.~{Jamrozy}, and S.~{Zola}, \aaa\ \textbf{371}, 445 (2001).

\bibitem{22:Amirkhanyan_n_en}
J.~{Machalski}, M.~{Jamrozy}, S.~{Zola}, and D.~{Koziel}, \aaa\ \textbf{454}, 85
  (2006).

\bibitem{23:Amirkhanyan_n_en}
K.~{Chy{\.z}y}, M.~{Jamrozy}, S.~J. {Kleinman}, et~al., Baltic Astronomy
  \textbf{14}, 358 (2005).

\bibitem{24:Amirkhanyan_n_en}
A.~{Buchalter}, D.~J.~{Helfand}, R.~H.~{Becker}, and
R.~L.~{White}, \apj\  \textbf{494}, 503 (1998).

\bibitem{25:Amirkhanyan_n_en}
L.~{Saripalli}, R.~W.~{Hunstead}, R.~{Subrahmanyan}, and
E.~{Boyce}, \aj\  \textbf{130}, 896 (2005).

\bibitem{26:Amirkhanyan_n_en}
V.~R.~{Amirkhanyan}, V.~L.~{Afanasiev}, and A.~V.~{Moiseev}, \ab
\textbf{70}, 45 (2015).

\bibitem{27:Amirkhanyan_n_en}
R.~M.~{Cutri}, M.~F.~{Skrutskie}, S.~{van Dyk}, et~al., VizieR
Online Data  Catalog \textbf{2246} (2003).

\bibitem{28:Amirkhanyan_n_en}
R.~M.~{Cutri}, et al., VizieR Online Data Catalog \textbf{2328}
(2014).

\end{thebibliography}
\end{document}